# Investigating Expectation Violations in Mobile Apps

Sherlock A. Licorish, Helen E. Owen, Bastin Tony Roy Savarimuthu and Priyanka Patel

Department of Information Science

University of Otago

PO Box 56, Dunedin 9054

New Zealand

{sherlock.licorish, helen.owen, tony.savarimuthu, priyanka.patel}@otago.ac.nz

## Abstract

*Information technology and software services are pervasive, occupying the centre of most aspects of contemporary societies. This has given rise to commonly expected norms and expectations around how such systems should work, appropriate penalties for violating these expectations, and more importantly, indicators of how to reduce the consequences of violations and sanctions. Evidence for expectation violations and ensuing sanctions exists in a range of portals used by individuals and groups to start new friendships, explore new ideas, and provide feedback for products and services. Therein lies insights that could lead to functional socio-technical systems, and general awareness and anticipations of human actions (and interactions) when using information technology and software services. However, limited previous work has examined such artifacts to provide these understandings. To contribute to such understandings and theoretical advancement we study expectation violations in mobile apps, considered among the most engaging socio-technical systems. We used content analysis and expectancy violation theory (EVT) and expectation confirmation theory (ECT) to explore the evidence and nature of sanctions in app reviews for a specific domain of apps. Our outcomes show that users respond to expectation violation with sanctions when their app does not work as anticipated, developers seem to target specific market niches when providing services in an app domain, and users within an app domain respond with similar sanctions. We contribute to the advancement of expectation violation theories, and we provide practical insights for the mobile app community.*

## 1. Introduction

Interactions between and among individuals in societies, both ancient and modern, have been governed by expectations [3, 4]. These expectations largely arise from societal norms and other rules or conventions of interaction that affect many facets of life. For example, long established gender-based role allocations have created anticipations (as well as tensions) about the nature of tasks those of a specific gender should undertake [5]. In the same vein, the norm of gift-giving at Christmas and birthday celebrations has been quite commonplace over a period of centuries. A more modern norm (or rule in some contexts) is seen in the growing implementation of *non-smoking practice in public places* of certain states. In fact, some initial norms are enforced, becoming rules, as in the *Equal Employment Opportunities (EEO) for everyone irrespective of one's gender or race* rule. Numerous behavioural expectations of this kind exist in ancient and modern societies [3, 4, 6, 7].

Such expectations are now encountered in the socio-technical domain, such as virtual communities and online platforms. On the basis that individuals are the main drivers of interactions in such systems, *expectations* regarding interaction behaviours of other entities, even in these more recently emergent contexts, naturally arise [8, 9]. This is fitting given the pervasive adoption and use of information technology (IT) in modern societies, and the associated globalisation and decentralisation of organisations (e.g., for military, commerce, agriculture, finance, education, communications, governments, private organisations, business bodies and the medical sector). This rapid adoption of IT mandates that interactions between globally distributed individuals and groups (both formal and informal) are increasingly mediated through information and communication technology (ICT) infrastructure.

Socio-technical systems are at the heart of such interactions [10]. Applications such as wikis, blogs, review portals and discussion boards are used by individuals and groups to start new friendships, explore new ideas, and provide feedback for products and services, instantiating socio-technical systems. In such systems it is common for communication traces from interactions to be archived, providing opportunities for researchers to gain insight into a range of emergent social, cultural and behavioural issues [11-13]. Of such issues, and given the way societies are governed by expectations, it is essential to learn about commonly expected norms and expectations, penalties for violating these, and more importantly, how to reduce the consequences of violations and sanctions. Explorations directed at addressing such issues could provide insights that would lead to functional socio-technical systems (i.e., with satisfied members), and general awareness and anticipations of human actions (and interactions) in this new paradigm [14]. Such outcomes could also extend current theoretical constructs and underpinnings [15, 16], and move the literature on expectation violation forward in these new contexts.

To contribute to such understandings and theoretical advancement we study expectation violations in mobile apps. In this domain, the construction of a technical system (an app[1]) is shaped extensively by the feedback provided by the users of the system [17-20]. Mobile apps have revolutionised the personal and social computing space wherein multitudes of apps, of which a significant proportion are free, are available to mobile device users. It seems no exaggeration to say that growth in the mobile app market has been phenomenal. The Google Play[2] store, the official and largest Android application repository, offers over 1 million apps to potential users, and the total number of app downloads has exceeded 50 billion [21], with similar statistics available for the alternative Apple App[3] store. The app community thus provides unique opportunities for research enquiries into users' expectations in socio-technical systems, and thus, we have exploited this opportunity in our portfolio of work. We have previously conducted preliminary work aimed at understanding if users of apps respond to expectations violations, where our outcomes confirmed that this was indeed the case. The preliminary outcomes are published in proceedings of the 2016 European Conference on Information Systems [14]. In this early study we used quantitative proxies (word usage patterns) to infer expectation violations. Here in the current work we go one step further to decompose this issue, using inductive techniques to explore both the nature of expectation violations and associated sanctions.

Using Expectancy Violation [1, 2, 15] and Expectation Confirmation [16] theories, we first studied how the mobile app community responds to expectation violations with sanctions when features do not work as anticipated. This enquiry provides the basis for our in-depth examination of sanctions given assumed app release strategy and the domain of consideration. On the basis of the ease of entering the mobile app marketplace and the many developers releasing apps, providers are likely to employ a policy aimed at perfecting the same features of other apps in the domain [22], or may deliberately target market differentiation based on service superiority [23]. While multiple apps providing the same service could be beneficial to users, this could also be problematic for the community, as sanctions may grow in relation to inadequacies of specific features and issues corresponding to the targeting of specific market niches (e.g. fitness tracking apps). We thus look to increment our initial enquiry and explore this issue here. Finally, generalised expectations can sometimes result in rules [24], and thus, users of specific apps will likely develop precise expectations around how particular features should be presented, thereby reacting with similar sanctions if these expectations are violated. Evidence of this pattern would have specific implications for both theory and practice (e.g., theory extension or revision of product release strategy). We thus explore users' reviews of three apps in the health and fitness domain to decompose these issues (refer to Section 2.3). Among our findings, we contribute to the advancement of Expectancy Violation and Expectation Confirmation theories, and we provide practical insights for the mobile app community.

[1] We use the term 'app' (or mobile system) to describe a program or software product that is frequently delivered (especially to mobile devices) via an online platform.

[2] https://play.google.com/store/apps?hl=en

[3] https://itunes.apple.com/en/genre/ios/id36?mt=8

In the next section we present the study's background and motivation, and also outline our specific research hypotheses (Section 2). We then describe our methodology, introducing our measures in this section (Section 3). Thereafter, we present our results (in Section 4), before discussing our findings and outlining their implications (in Section 5). We then consider the threats to our study in Section 6, before finally providing concluding remarks in Section 7.

## 2. The Study of Expectations

Expectations are beliefs about future events and/or outcomes (i.e., a belief that something will happen or is likely to happen[4]). Individuals hold a range of expectations regarding the behaviour of others [3, 4]. In this regard, expectation as a concept has been studied from various perspectives, including norms, conventions, policies and pledges [9]. Although possessing subtle differences depending on the field of study, generally, when an expectation is not fulfilled (or, is violated) those affected become dissatisfied, and their reactions may be expressed – perhaps publicly – in some form of resentment or reprisal. Expectation violations could thus result in an external behavioural change (e.g., a verbal or material sanction) or an internal record of violation (e.g., loss of reputation) [1, 2]. For example, a research advisor expects her students to be on time for their research meetings, and a chairwoman expects her board members to be prepared for their monthly meetings. Violation of either expectation/norm would result in some form of reaction or sanction (e.g., a conversation enforcing expectation or written warning).

There has been substantial research effort aimed at understanding various forms of human-human expectations in the social sciences [3, 4, 6, 7]. For instance, in the examination of the way social norms and expectations are formed, of the forces determining their content, and of the cognitive and emotional requirements that enable individuals to establish and enforce social norms, it was observed that sanctions are decisive for norm and expectation enforcement, and that they are largely driven by non-selfish motives [7]. In addition, Fehr and Fischbacher [7] noted that studying social norms and expectations provides insights into the proximate and ultimate forces behind human cooperation. They also established that cooperation is achieved voluntarily or through sanctions. Oliver and Burke [25] studied the role and persistence of expectation and expectation-related effects within the expectancy disconfirmation and performance model (also known as Expectation Disconfirmation Theory (EDT)) and found that expectation-initiated performance comparisons (disconfirmation) and performance judgments were important satisfaction influences. In particular, these authors showed that expectation manipulation had an immediate but declining effect on consumption of a product or a service. Researchers have examined this issue from other perspectives, including for example, from a normative viewpoint [6] and that of culture adherence [4].

While some attention has been directed to the study of the fulfilments and violations of expectations that are expressed in socio-technical systems and ICT-mediated interactions [26-29], attention has not been given to expectation fulfilment in the mobile app space. In fact, prior work on computer-mediated expectations exists largely in the area of human-computer interaction (HCI), where the focus has been on designing user interfaces that meet the expectations of intended users of conventional software systems (e.g., windows-based applications) [26, 30]. However, a growing range of data sources are now available which contain evidence of users responses to expectations and their fulfilment or violation in a range of other socio-technical domains, including the mobile app community [31]. The hidden responses to expectations (and their violations) in these sources can be extracted and analysed to obtain valuable insights. Recent works have indeed noted that large numbers of requests for software improvements are logged by online communities [17, 19, 20, 32]. These requests are likely to be fuelled in part by stakeholders' expectations about how the mobile software (e.g., apps) should function (i.e., human-software expectations). Violations of expectations, in particular, are important to study as they point towards potential reparations that may result if stakeholders' expectations are continually violated. In addition, given that feedback (including requests) logged online are not mediated by the pressures of participant observation or socially desirable responding, the data captured by online repositories are noteworthy for studying expectation violations and the reactions they elicit. Beyond practical implications, such insights would also advance expectation theories. We consider this issue next. Thereafter, we review the way expectation violations have been studied in socio-technical systems, before developing our hypotheses and study model.

[4] http://www.merriam-webster.com/dictionary/expectation

### *2.1. Expectation Violation*

As noted in the previous section, the issue of expectation violation has generated considerable research interest [3, 4, 6, 7]. In the specific context of human communication, the Expectancy Violation Theory (EVT) has been produced [1, 2, 15]. EVT notes that individuals form expectations based on societal rules which drives their beliefs about future events, and should the expectation of an individual be violated, they react in compensation [8, 33]. For example, in the field of proxemics, if someone stands too close while having a conversation with another person, the second individual may feel uncomfortable, changing their gesture and posture [27, 34, 35]. This is because their expectation of the inter-personal distance, arising from a culture-specific norm, has been violated during the conversation. Moving beyond non-verbal cues, among other considerations, EVT has been studied in the context of verbal communication, e.g., teamwork in organisations [15] and higher education [36]. In these and other disciplines there has been support for EVT in explaining individuals' behaviours and responses (including adherence) to expectations and norms.

Similar to EVT, conventions, norms, contracts, policies, trust, pledges and commitments also drive vendor-purchaser behaviours, and these concepts are thus embedded in consumer satisfaction theories such as the Expectation Disconfirmation Theory (EDT). EDT, also known as Expectation Confirmation Theory (ECT), was initially proposed in the marketing domain [16]. ECT notes that post-purchase fulfilment is influenced by previous expectations, perceived performance, and disconfirmation of beliefs. This theory has been successfully employed to study the role of customer expectations on the adoption, usage and continuity of services [37, 38]. Such actions are quite related to those that are likely to be prevalent among mobile app users.

In general then, ECT and EVT could provide useful bases for exploring expectations fulfilment or violation in socio-technical domains. In modern societies and globally connected organisations where ICT enables the instant sharing of information, violations of expectations and sanctions may be reported in a range of social media outlets such as social networks, discussion and review boards and emails [39, 40]. Thus, extending the predominantly human-human notion of expectations to the domain of human-software (and its use) would provide insights into how individuals and groups have extended this concept to the requirement of software predictability during use [41, 42]. We examine this issue further in the following section.

### *2.2. Expectation Violation in Socio-technical Systems*

The socio-technical domain comprises humans using applications and/or software entities to achieve specific goals (whether organisation or personal). With the rapid adoption and use of ICT in modern society, ICT-mediated interactions embody expectations [8, 9]. To this end, humans have naturally extended the notion of expectation to the functioning of software [41, 42]. They expect software to behave in a certain way, particularly if they have had previous experience with similar software (e.g., they would likely expect similar menu conventions from software products belonging to the same vendor). Individuals' expectations may also arise due to recommendations from others, or from the reading of product reviews [26, 30]. These expectations give rise to beliefs, and anticipated behaviours of 'software entities'. Violations of those beliefs have implications for user satisfaction, and continuation of software use, as posited by ECT. While researchers have studied users' expectations of different types of information systems [43-46], the nature of sanctions arising as a result of expectation violation of information systems use has not been scrutinized.

Specific expectations may also arise through the ubiquitous usage of mobile apps across domains, including life logging, social networking, banking and health (among others). As a result, there is increasing calls towards providing insights into the nature of expectation violations (and sanctions) faced by individuals progressively using apps to address personal and work obligations [19, 20, 32]. Questions such as *what are the expectations for particular mobile systems?*, *what are the penalties or sanctions for violating expectations?*, and more importantly, *how can consequences to violations be reduced?*, offer an avenue to enhance our understanding of these systems and how they shape societies. It would thus be insightful to explore this phenomenon towards answering these questions. As noted above, a prominent source from which such knowledge may be drawn is the feedback portal that is commonly provided to app users. Information conveyed on such portals is publicly available, providing a potentially rich and valuable source of details regarding stakeholders' expectations, how these are (or are not) accommodated, and the penalties or sanctions that expectation violations evoke. We framed three hypotheses to address such an enquiry, as outlined next.

### *2.3. Hypotheses Development*

While the consideration of EVT [15] and ECT [16] has to date been limited to human-human expectation violations, we contend here that such theories may be applied to the study of expectation violations in socio-technical systems, and the mobile app community in particular [14]. Users must increasingly interact with software in contemporary societies, giving rise to expectations about how various types of software 'should behave'. As with the way humans develop expectations about specific societal conventions and norms, which are driven by an innate system of reasoning [3, 4, 6, 7], so too are individuals likely to develop expectations about how specific software should behave [8, 33]. In human to human context, when an individual's expectation is not fulfilled (or, is violated) they become disappointed, and their reactions are sometimes expressed in the form of sanctions. Such sanctions may be publicly expressed, or articulated in subtle forms of resentment [1, 2]. This phenomenon is likely to be evident among members of the mobile app community whenever there are violations to their expectations about how features and services should operate. Given the many channels now available (e.g., social networks, wikis, blogs, review portals and discussion boards) to members of mobile app communities to express such resentment, responses or sanctions (i.e., reactions as a form of compensation [8, 33]) in relation to expectation violations could be numerous. In fact, there is scope to extend EVT and ECT; for instance, while ECT explains that customers' fulfilment is linked to their previous expectations, perceived performance, and disconfirmation of beliefs, little is explored around how sanctions may result through such violations. In providing insights into this issue and validating the relevance of EVT and ECT to a specific socio-technical instance, that of mobile apps, we outline our first hypothesis.

> ***H1**. A mobile app community responds to expectation violations with sanctions in their reviews when features do not work as anticipated.*

With the ease of entering the mobile app marketplace and the many developers releasing apps, the need to be nimble means that apps are developed with the view of perfecting the same features of other apps in the domain (e.g., fitness tracking apps) [20, 47, 48]. This approach is referred to as a "non-additivity" strategy among those that considered the demand theory [22], where particular market niches[5] are identified based on the inadequacy of others' services. From a marketing perspective, this trend follows Porter's strategy aimed at market differentiation where products and services are qualified based on their superiority [23]. In fact, the approach to adopt a non-additivity strategy may also find favour among users that are driven by features they are familiar with, especially having used similar apps that performed unsatisfactorily. Studies across various disciplines have shown this effect, where what is popular tend to be given priority [49, 50], in a way fuelling the notion that "popularity begets popularity" [51]. We aim to validate these effects through our second hypothesis.

> ***H2**. Sanctions in relation to expectation violations are about specific features given app developers' drive to target specific market niches.*

Societal norms take varying forms (e.g., consumption norms, behavioural norms, and norms of cooperation [52]), and often influence individuals to develop context-specific expectations, also termed generalised expectations [53]. These generalised expectations can sometimes influence formal rules or laws [24]. Violation of such expectations can thus lead to various degrees of sanctions or punishments, from being shunned by members of society (e.g., for not paying a tip after dining at a restaurant) to being prosecuted in a court of law (e.g., for smoking in public spaces in some countries). Cultural factors in particular are said to have a significant influence on most norms, and how these vary across social contexts. For instance, maintaining eye contact while communicating is perceived positively by North Americans with European ancestry [54], while the opposite is held for Novaho Indians residing there [55], as is the case for Japanese and Ethiopians [56, 57]. In other manifestations of the cultural effects on societal norms, while North Americans perceive those that look down while communicating to be lacking knowledge or confidence, this is seen as a symbol of respect among Japanese [57]. Similar divergence is seen for expectations regarding maintaining inter-personal distance for different cultures [27, 34, 35].

[5] Product features aimed at satisfying specific market needs.

We anticipate that this would transcend to individual expectations in socio-technical systems. Users of specific apps will likely develop specific expectations around how particular features should be presented, and thus, are likely to react with sanctions if these expectations are violated. While the literature has accepted the notion that actors in socio-technical systems are likely to develop expectations around the workings of software [8, 9], previous work has not actually looked to formally test this conjecture. Insights into these reactions could be useful for understanding, and planning for, human behaviour in mobile app community. In addition, such insights may have implications for validating and extending EVT and ECT theories. We thus developed a third hypothesis to examine this phenomenon.

> ***H3***. *Users of a specific app domain will respond with similar sanctions to expectation violations given their anticipations of how software features should work in that domain.*

These three hypotheses (H1-H3) are depicted in the model in Figure 1. We first explore whether expectation violations induce sanctions from the mobile app community when app features do not work as anticipated (H1). Thereafter, we investigate whether such sanctions are often in relation to specific features, given developers' drive to target specific market niches (H2). Finally, similar to the way culture affects expectations and norms (noted above), we examine whether app users' sanctions are driven by domain expectations (H3). We present our methodology in the next section, comprising an outline of our data source and operationalisation of our variables.

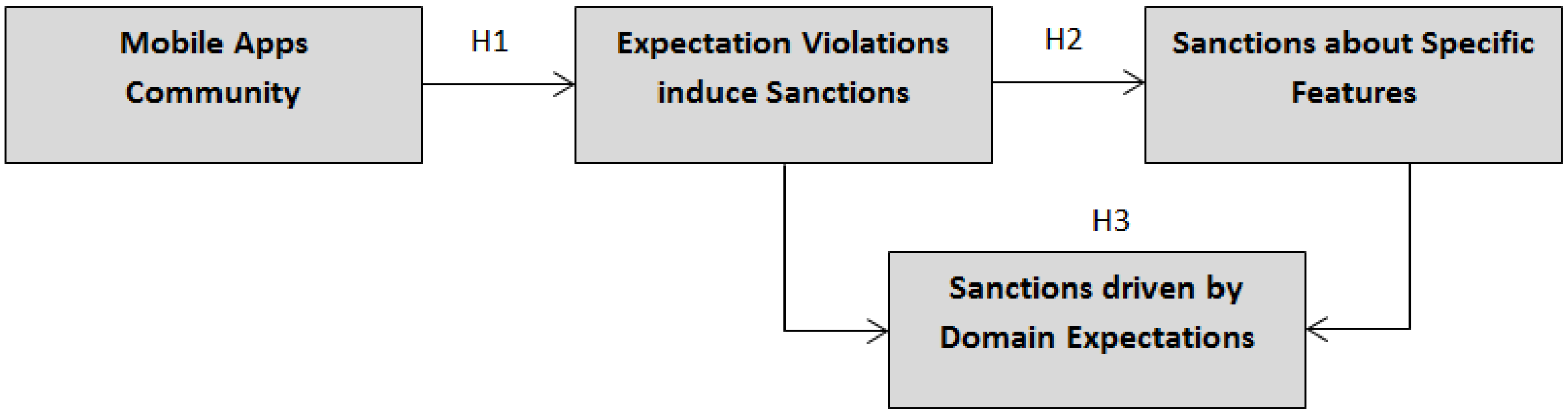

*Figure 1. Study model*

## 3. Methodology

In light of the research issues presented in the previous section this study utilises a mixed method approach [58]. This approach is implemented under a case study design, taking on both confirmatory and exploratory tones [59]. According to Yin [60], case studies are used to investigate contemporary issues in real settings. In particular, Yin posited that this method is generally suitable when there are unclear boundaries between phenomena and context [60], as is the case in this work (refer to previous section). Thus, this method provides an avenue to understand *what are the expectations for particular mobile systems*, *what are the penalties or sanctions for violating expectations* and *how can consequences to violations may be reduced*, towards enhancing our understanding of these systems and how they shape societies [61]. Exploratory aspects of the case study are used to initiate or extend theories, while previous theories are assessed using more confirmatory aspects [62]. We note that although we intend to test three hypotheses (H1-H3), the relevance of the theories for unpacking the issues under consideration is unclear. The work thus demands exploratory and confirmatory tones.

A case study may employ purposive sampling so that relevant cases are selected for observation [60]. Sometimes the most representative cases are selected, but abnormal cases may also provide interesting observations [61]. Such a mixed approach is used during this work's case selection process, and deliberate efforts are undertaken to ensure that interesting variations in mobile apps are captured during data sampling. In fact, case study research employing multiple cases provides strong claims for validity [62], we thus adopt this approach.

The unit of analysis provides the basis for how data is collected and analysed, whether at the company level, project level, team level or individual level [62]. We use multiple units of analysis, at the community (multiple companies) and app levels (company) [63]. The study is thus conducted using a multi-phase approach, first employing exploratory analysis techniques (content analysis), before more quantitative statistical techniques are used for further analysis. Since this study uses archival data, data mining principles are used for data collection, pre-processing and preliminary data exploration. Extracted data are then analysed using content analysis and statistical techniques for testing H1, H2 and H3. As a part of our preliminary data exploration we have also used sociograms [64] to model and visualize various relationships between concepts (e.g., features and sanctions). We provide further details around our dataset, techniques and measures in the three subsections below.

### *3.1. Data Collection and Pre-processing*

To facilitate our inquiry we extracted reviews from Google Play, using the android-market-api[6]. This API allows the extraction of reviews (4,500 most recent) that are logged by users for apps on Google Play. The reviews comprising author ID, creation time, rating and comment text, were extracted using a script written by the last author of the paper between March 2015 and February 2016. Thereafter, the reviews were pre-processed by removing punctuation, tags and non-English characters so as to avoid confounding our analysis [65]. The cleaned reviews were then stored in a database, where exploratory data analysis involving querying the structure of the reviews and assigning the most suitable datatypes to specific fields was conducted. Through these exercises we noted that reviews sometimes discussed multiple issues in different sentences. However, given the goal of our study, we anticipated that partitioning each review into individual sentences would compromise the overall thoughts expressed by users. We thus used the review as our unit of analysis to capture users' sanctions, and isolated aspects that were not related to expectation violations during our analysis (e.g., praise or positive sentiments).

We extracted 12,419 reviews from three apps (MyTracks[7], Endomondo[8] and Zombies Run![9]) belonging to the health and fitness domain. The three apps were chosen based on their popularity in the domain over the past years. The popularity score was based primarily on usage statistics and rankings (e.g., all three apps have one million or more installations), which were both available through the use of the Google's android-market-api (described above). Similar numbers of reviews were logged for two of the apps (MyTracks=4471 and Endomondo=4442), with the last app attracting fewer reviews (Zombies Run!=3506). The number of sentences in the reviews (or length) were also very similar for the first two apps (MyTracks=8491 and Endomondo=8623), with users logging slightly shorter reviews for 'Zombies Run!' (5609 sentences). This data was suitable for conducting our planned analyses.

We anticipated that responses to expectation violations would possess multiple attributes, and this was indeed confirmed by our previous work [14]. We first filtered reviews that attracted poor ratings (between 1 and 3 out of 5). This step was motivated by Fu et al. [66], who observed that reviews that were rated =< 3 indeed largely contained expressions of users' dissatisfaction, though, the nature of violations was not considered by these authors. Of that subset of reviews we next extracted those that contained negative words (i.e., either negative emotion words or generic negative words). Negative words, such as *dislike*, *sad*, and *hate*, have been observed to signal individuals' dissatisfaction [67]. Thus, in order to support our analysis such words were obtained from the Linguistic Inquiry and Word Count (LIWC) library [67]. The LIWC tool library has been developed by researchers over the past 30 years and has been widely used in similar work; e.g., [68]. A total of 431 negative words were extracted to support our analyses. For a full list of such words, we refer the reader to LIWC online[10]. Having selected our bag of negative words (431 in total), we tokenized the low-ranked reviews (i.e., rated =< 3) into words, and then employed a pattern matching approach to extract those that contained expressions of dissatisfaction. If the review contained at least one negative word it was classified as an expression of dissatisfaction (implying there was a potential response to some expectation violation).

[6] https://code.google.com/p/android-market-api/

[7] https://play.google.com/store/apps/details?id=com.zihua.android.mytracks&hl=en

[8] https://play.google.com/store/apps/details?id=com.endomondo.android&hl=en

[9] https://play.google.com/store/apps/details?id=com.sixtostart.zombiesrunclient&hl=en

We anticipated that, beyond the LIWC tool's negative words (e.g., *sad* and *dislike*), other words and phrases may also combine to indicate dissatisfaction (e.g., *does not*, *cannot* and *must not*). We indeed observed this during our preliminary exploratory analysis. Thus, another bag of words containing 57 such phrases, including negative modal verbs (e.g., *must not*, *should not*, *may not*) and common slangs such as "*ain't*", and "*shan't*" was constructed. This way, even if a review did not contain negative words as captured by the LIWC dictionary, it would still be labelled as a violation review if it contained one or more of the 57 additional phrases in the second bag of words. In perusing the literature to further understand word use and its relationship to the study of norms and expectations, we noted that modal verbs (e.g., *must* and *should*) have been used previously to identify obligations in emails and business contracts [69, 70]. Negations also tend to correspond to expectation violations arising from obligations and prohibitions [71]. Obligations describe what someone is expected to do, while prohibitions describe what individuals are not expected to do [72]. We anticipated that evidence of responses to both forms of violation may exist in app reviews, and thus, our inclusion of the negations of such verbs (e.g., *must not*) would indeed capture additional responses to violations beyond those extracted through the use of the LIWC negative words. Of the 12,419 total reviews, 2062 (My Tracks=819, Endomondo=672, Zombies Run!=571) were classified as containing responses to expectation violations – in the form of sanctions (i.e., close to 17% of the reviews). The above process is illustrated using a sample review in Figure 2.

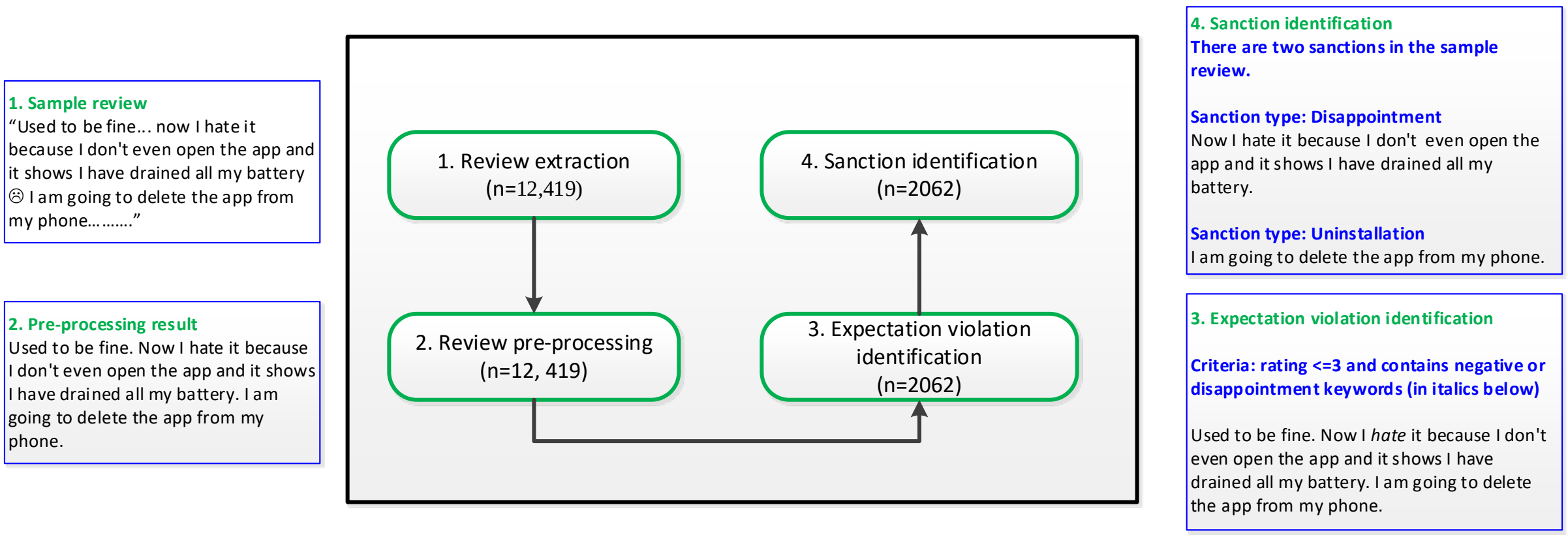

*Figure 2. The process employed for data collection, pre-processing and sanction identification*

To initially validate that these reviews indeed expressed users' dissatisfaction as a result of expectation violations, the first and last authors coded 400 randomly selected violation sentences from a total of 4525 sentences (from the 2062 reviews). These were labelled with Y or N depending on if they were assessed as representing dissatisfaction as a result of a violation or not. We found that only 17 sentences were wrongly selected, indicating a 96% agreement. We are thus confident that the approach used above indeed reliably selected reviews that contained expressions of dissatisfaction as a result of expectation violations. That said, overall, some of the reviews also contained positive appraisal of some app features (in addition to the negative aspects), and thus, we conducted another round of reliability checks after thematic coding the entire dataset. This and additional steps taken during the analysis are discussed in the following subsections, after our terms and measures.

[10] http://www.liwc.net/

### *3.2. Terms and Measures (H1-H3)*

*Responses to expectation violations*: Consistent with Expectancy Violation [1, 2, 15] and Expectation Confirmation [16] theories, we defined *expectation violation* of mobile fitness apps as the perception that app features do not work as anticipated. This was explored across three mobile apps: MyTracks, Endomondo and Zombies Run!

*Sanctions*: Sanctions here are defined as users' resentments expressed in the form of complaints, decisions, reactions and warnings [1, 2]. We measured app users' reactions to expectation violation by the prevalence of particular *sanctions*.

*Features*: Expectation violations of mobile apps were determined by the type and frequency of domain-specific problems that were associated with app features. We characterized app *features* by specific components (e.g., GPS, battery, help, subscription) that were faulty, affecting app usability. Problems (*Issues*) were identified by cases in which the user experienced a feature malfunction; for instance, excessive battery *drainage*.

*Mobile app community or specific app domain*: MyTracks, Endomondo and Zombies Run! are considered to be in a specific app domain because they are socio-technical systems that operate independently of each other and allow the user to carry out similar fitness regimes. These apps and the users that utilise them make up the mobile app community.

We consider how these measures are computed during our analysis in the remaining subsections.

### *3.3. Content Analysis Approach*

Holsti [73] explained that content analysis (CA) may be used to describe trends in communication content, relate known characteristics to a source, and compare content to standards. Fundamentally, content analysts use communication in seeking answers to questions such as "who says what, to whom, why, to what extent and with what effect". This technique was developed by Alfred Lindesmith in 1931 to provide an alternative view for exploring, evaluating and analysing meaning, besides testing hypotheses, and was later supported by Glaser and Strauss in the 1960s in their variant of this methodology known as Grounded Theory [74].

Under the qualitative content analysis umbrella, analysis techniques range from intuitive and interpretive analyses to the more strict examination of data [75]. The specific approach chosen for implementation is often aligned with the researcher's theoretical stance and the nature of the phenomenon of interest [76]. Hsieh & Shannon [77] classify these approaches as conventional, directed and summative. The conventional form of CA is used to describe phenomena where existing theories are limited, thus, the approach is aimed at theory building. As with the implementation of the grounded theory method, researchers employing this approach generally start the process of data analysis by inductively examining the data, allowing meaning to flow from the data as against approaching data analysis with any preconceptions [78].

Given the pragmatic approach adopted to test our hypothesis (H1-H3), we employed a more conventional form of CA, where we inductively [79] explored the data to see if clear sanctions were evident in app reviews (to test H1). The first three authors each examined 20 reviews for sanctions, where sanctions were indicated if user expressed disappointment or threatened the app owner with a penalty for violating some form of expectation (e.g., "I have decided to use App X over this one, as the distance metrics on this app are all incorrect."). We observed that there was indeed evidence of various forms of sanctions, from turning to competition, warning to users, to disappointment. We thus decided to follow the established recommendation for formal coding [78], where each review was initially read through for familiarization, and codes were then assigned to specific aspects of the review that denoted a sanction in response to a violation.

This process allowed us to measure the prevalence of specific feature malfunctions and their resulting sanctions. Initial codes for features and issues were identified and extracted based on explicit, surface-level semantics in the data (see [80]). Then, we used thematic mapping to restructure the codes in terms of sanction types, which served as the overall themes.

We started this process with MyTracks, and coded the 819 reviews (filtered for sanctions, noted above) in developing eight themes of sanctions (refer to Codes 1 to 8 in Table 1). Thereafter, we coded the Endomondo reviews (672 in total) where a ninth sanction (theme) emerged (refer to Code 9 in Table 1). Two additional sanctions then emerged after coding the 571 Zombies Run! reviews (refer to Codes 10 and 11 in Table 1). This evolution of codes is normal when existing theories are insufficient to capture all relevant insights during preliminary CA coding [77]. For each phase of the coding we conducted a form of reliability checking where 20 reviews were randomly coded by the first three authors to ensure that there was similar interpretation of the sanctions. Initial checks confirmed between 86-90% agreement, where further discussion always led to 100% convergence. The minor differences noted largely resulted due to differences in the number of codes that were assigned to some sanctions.

To this end, we agreed that multiple codes should be assigned to utterances that demonstrated more than one form of sanctions. We conducted another round of checks at the end of coding exercise and achieved 90% inter-rater agreement between the first three authors as measured using Holsti's coefficient of reliability measurement (C.R.) [73]. This represents excellent agreement between coders and suggests that a consistent and reliable approach was taken. Codes were analysed using formal statistical methods, however, given the large dataset we also use sociograms developed using social network analysis (SNA) for preliminary exploration of the relationship among codes. These techniques are examined next.

*Table 1. Sanctions in response to expectation violation across app*

| Code | App | Sanction | Examples |
|---|---|---|---|
| 1 | All apps | Disappointment | It just does not work as it used to.<br>Hoping it might have improved significantly…really disappointed.<br>Such a bad game. |
| 2 | All apps | Complaints | Please fix this app…<br>Google, you can do better than this.<br>Don't have time to sit here waiting for the app to work. |
| 3 | All apps | Turn to competitors | Mototracktour tracked my route perfectly –whilst this app lost about 30 of the route<br>Try sportstrackerlive instead |
| 4 | All apps | Giving or seeking advice | Would anyone be able to tell me how I can….<br>It would be nice to be able to…<br>It would get a five star rating if this issue is fixed. |
| 5 | All apps | Uninstalled | Getting worse and worse, removed<br>I won't be using it again until these are fixed. |
| 6 | All apps | Warning to users | Don't buy until this is fixed.<br>Do not download…<br>Don't recommend if avid hiker. |
| 7 | All apps | Abuse | This is f***ed up.<br>You're getting more than $30 worth of an angry user who feels like he was ripped off. |
| 8 | All apps<br>(Removed) | False positives | I like it how easy it is to….<br>I like it as it is… |
| 9 | Endomondo<br>Zombies Run | Payment complaints | Shame I paid for year as a premium member.<br>Can't even start playing without being redirected to buy something. |
| | | | I feel ripped off. A cynically overpriced app. |
| 10 | Zombies Run | User blames self | Not what I thought, but that's my fault.<br>I could never figure out how to get it to work. |
| 11 | Zombies Run | Embarrassment | I'm not running around holding my cell in front of me.<br>This is a game for people that don't have a life. |

### *3.4. Social Network Analysis*

We explore the relationship among codes and pronounced sanctions about specific features and problems (issues) using Social Network Analysis (SNA). This helped us to explore the similarities between sanctions for specific feature-issue associations. SNA is used to quantify aspects of network structures in order to support pattern identification among related items [81]. This technique employs mathematical analysis and pictorial representations of the patterns of interaction and relationships among group processes [82]. Concepts such as cohesion, equivalence, power and brokerage are used to explain the characteristics of network nodes [83]. Visualization of interaction networks, also called sociograms, is often used for uncovering interconnections and the flow of information that may not be so evident from numerical values [64]. In these visualizations, elements are represented by nodes, and their associations are illustrated through lines that connect these nodes. An examination of a sociogram will quickly unveil pronounced items; refer to Figure 3 for illustration. Here the graph is configured to represent features (e.g., GPS, Battery, Track) using black circles. The larger circles represent features that received a higher numbers of sanctions. Edges (lines) are modelled to be thicker for feature-issue pairs that appeared more frequently (e.g., GPS-Wrong, Battery-Drain).

SNA has been shown to be valuable in many domains, including security [84], political science [85], education and communication [86]. SNA techniques also provided utility for this work; visualisations were used to study pronounced sanctions about specific features and issues.

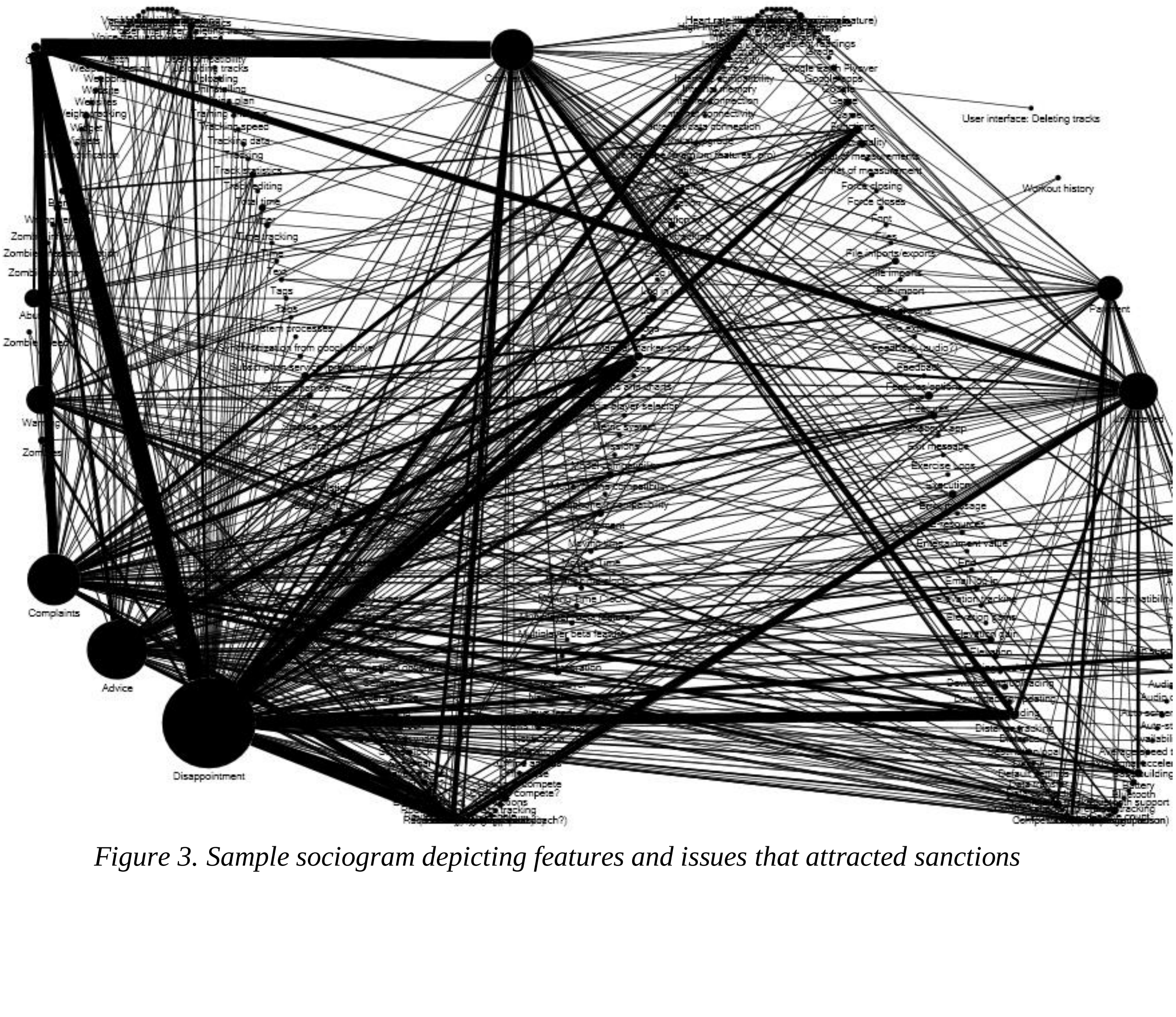

*Figure 3. Sample sociogram depicting features and issues that attracted sanctions*

## 4. Results

We firstly tested whether app users expressed resentment to expectation violations through the use of particular sanctions. After eliminating two unclear cases of sanctions, we calculated the overall frequencies of different sanctions across MyTracks, Endomondo and Zombies Run!. A total of 2,948 sanctions were recorded in response to expectation violation. We further eliminated 446 false positive sanctions (coded 8) in which the users expressed satisfaction about particular features of apps, and three misclassifications of abuse. Thus, our final sample consisted of 2,502 sanctions. Through the thematic mapping process, ten groups of sanctions were extracted, a summary of these is provided in Table 2. We use our outcomes to test the three hypotheses (H1-H3) outlined in Section 2, first providing our outcomes in relation to H1 in Section 4.1. We then provide results to test H2 in Section 4.2, before providing outcomes for testing H3 in Section 4.3.

### *4.1. App community responds to expectation violations with sanctions (H1)*

Aggregating the sanctions across all the three apps in the health and fitness domain, we calculated the prevalence of sanction types in response to expectation violations. As shown in Table 2, fitness app users were most likely to respond to expectation violation through expressions of disappointment. They also responded frequently to expectation violation by complaints, turning to competitors, advice giving and seeking and uninstalling. Our evidence here offers support for H1, that **a mobile app community responds to expectation violations with sanctions in their reviews when features do not work as anticipated.**

*Table 2. Frequency of sanctions in response to expectation violation across the three apps*

| Sanction | Count | % |
|---|---|---|
| Disappointment | 892 | 35.7 |
| Complaints | 404 | 16.1 |
| Turn to competitors | 338 | 13.5 |
| Advice giving or seeking | 279 | 11.2 |
| Uninstalling | 248 | 9.9 |
| Warning to users | 158 | 6.3 |
| Payment complaints | 100 | 4.0 |
| Abuse | 58 | 2.3 |
| User blames self | 15 | 0.6 |
| Embarrassment | 10 | 0.4 |

### *4.2. Expectation violations about specific features (H2)*

**Features**: we conducted chi-squared tests for significant differences in the prevalence of sanctions in relation to features. Taking the 10 most frequently occurring features across the top five sanctions, we found that expectation violations invoked sanctions frequently for 20 unique features (see Table 3). For our analyses, we excluded a further 8 features with expected counts of less than five across the sanctions as the test statistics are too deviant for chi-squares [87]. We also excluded one feature for which there were no complaints associated with a particular sanction. There was a significant association between features and sanctions, $\chi^2(40)$ =325.07, Cramer's $V$ = .26, $p$ < .001.

As shown in Table 3, disappointment sanctions were most often expressed towards the game Zombies Run! (rather than towards specific features of the game) and were significantly more prevalent than expected by chance, $z$ = 7.1, $p$ < .01. Almost 12% of expressions of disappointment were directed at the Zombies Run! app. At the feature level of analysis, notwithstanding our analysis in relation the top five sanctions, 86.4% of Zombies Run! malfunctions were responded to with expressions of disappointment. In terms of individual app features, users responded most often to GPS failure, followed by recording, maps and distance tracking malfunctions. However, despite this, disappointment was expressed less often towards GPS functions than expected, $z$ = -2.7, $p$ < .01. Only 32.6% of GPS malfunctions were associated with expressions of disappointment[11]. In contrast with other sanctions, users responded significantly less often to privacy violation (e.g., Google tracking personal calls and phone contacts) with expressions of disappointment, $z$ = -4.7, $p < .01$, where disappointment was only expressed in response to 3.3% of cases of privacy violation.

[11] We only calculated the percentages of the top five sanction in responses to expectations associated with a particular feature.

Complaints to app proprietors were again most frequently expressed towards GPS as a single feature. This was followed by maps, distance tracking, privacy violation and installing/updating, although the counts were smaller. Users also directed their complaints at MyTracks app without targeting specific features. Even though privacy violation and installing/updating accounted for only 5% and 4.5% of total complaints respectively, only these features were associated with significantly more complaints than expected, $z_{\text{privacy violation}} = 3.0$, $p < .01$, $z_{\text{installing/updating}} = 2.5$, $p < .05$. At the feature level, 32.8% of privacy violation problems and 30% of installing/updating malfunctions were responded to with complaints. In contrast with disappointment, complaints were significantly less likely to be directed at the Zombies Run! app, $z$ = -3.6, $p < .01$.

GPS malfunctions accounted for 27.2% of instances of users turning to competitors. At the feature level, 31.9% of GPS malfunctions were responded to by users turning to competitors; an association that occurred significantly more often than expected, $z$ = 5.5, $p < .01$. Users also turned to competitors in response to distance tracking, recording, maps and route tracking feature malfunctions; however, none of these features lead users to turn to competitors more often than expected. Users turned to competitors significantly less often than expected because of malfunctions with audio, $z$ = -2.1, $p <$ .05, and installing/updating, $z$ = -2.7, $p < .01$. Users also turned to competitors less often than expected in relation to general Zombies Run! app malfunctions, $z$ = -4.4, $p < .01$.

Malfunctions with maps accounted for 13.3% of expressed advice seeking or giving. At the feature level, users responded to 24.3% of map malfunctions with advice-based sanctions, which was significantly more often than expected, $z$ = 5.8, $p < .01$. Users also responded frequently with advice-based sanctions towards GPS, but not more often than expected. However, users gave and sought advice more often than expected about phone compatibility, $z$ = 2.7, $p < .01$, and audio malfunctions, $z$ = 2.6, $p < .01$. Interestingly, Zombies Run! was less likely to induce advice-based sanctions than expected, $z$ = - 2.8, $p < .01$.

Finally, sanctions expressing a desire to uninstall the app were communicated more in response to GPS failure and privacy violation, accounting for 14.5% and 11.7% of sanctions respectively. At the feature level, uninstalling sanctions accounted for 54.7% of privacy violation, which was significantly higher than expected, $z$ = 8.0, $p$ <. 01. Uninstalling apps was also associated with distance tracking and recording problems; however, these problems were not associated with more threats and decisions to uninstall.

**Issues (Problems):** Again, we conducted chi-squared tests to examine significant differences in the prevalence of issues that resulted in sanctions. Taking the ten most frequently occurring issues per sanction, we found that expectation violation was expressed in response to 20 different issues across the five sanctions (see Table 4).

We conducted two stages of preliminary chi-squared tests to eliminate all the issues with expected counts of less than 5. However, 17 out of the 20 issues had expected counts of less than 5. Therefore, we conducted a final chi-squared test with only 3 issues for which expectation violation was frequently expressed across sanctions: inaccurate tracking, GPS signal loss and the perception that the app was bad/boring. The prevalence of these issues differed significantly across sanctions, $\chi^2(8)$ = 101.46, $p < .001$, Cramer's $V$ = .30.

As shown in Table 4, disappointment sanctions were most frequently expressed when the user believed the app was bad/boring, an issue which accounted for 14.9% of disappointment sanctions. Users responded to 77.3% of bad/boring perceptions with expressions of disappointment, which was significantly more often than expected, $z$ = 5.4, $p < .01$. However, disappointment sanctions were expressed significantly less often than expected towards inaccurate tracking, $z$ = -2.5, $p < .05$, and GPS signal loss, $z$ = -2.7, $p < .01$, which collectively only accounted for 15.8% of disappointment sanctions. At the issue level, users responded to 38.1% of inaccurate tracking and 31.4% of GPS signal loss with sanctions of disappointment.

Complaints were most frequently expressed in response to inaccurate tracking, an issue that accounted for almost 13% of total complaints. These complaints were expressed more often than expected, $z$ =

2.4, $p < .05$. However, on the whole, users directed their complaints at a variety of issues rather than targeting a specific set, as indicated by the smaller counts in Table 4. GPS signal loss was not associated with complaints beyond chance. Surprisingly, perception that the app was bad/boring was associated with significantly fewer complaints than expected, $z = -2.0$, $p < .05$.

Turning to competitors was also most frequently expressed towards inaccurate tracking and GPS signal loss; these issues accounted for 18.6% and 14.2% of turning to competitor sanctions respectively. However, at the feature level, users responded to 39.7% of instances of GPS signal loss issues by turning to competitors, which was significantly more often than expected, $z = 4.3$, $p < .01$. However, apart from these most prevalent issues, sanctions were expressed evenly across a variety of issues. Users turned to competitors significantly less often than expected on the basis of the app being bad/boring, $z = -4.3$, $p < .01$.

Advice-giving and -seeking were not directed at specific issues. Consistent with the other sanctions, users responded most frequently to inaccurate tracking with advice-giving sanctions. However, users did not respond to inaccurate tracking with advice-based sanctions beyond chance expectancy, and the other issues yielded such small counts they could not be included in the chi-squared tests. Consistent with turning to competitors and complaints, advice-based sanctions were negatively associated with the perception that the app was bad/boring, $z = -2.3$, $p < .01$.

Finally, threats and decisions to uninstall were most frequently expressed in response to tracking inaccuracy (13.3%). GPS loss and general dissatisfaction with the apps also resulted in threats and decisions to uninstall, but at a lower rate. However, none of these issues were significantly associated with uninstalling beyond chance. Although some sanctions were expressed towards a variety of features and issues (resulting in low frequency counts), the results show that users' responses to expectation violation are largely about specific features (i.e., GPS, maps, distance tracking and recording) and issues (i.e., inaccurate tracking, GPS signal loss). This supports H2, and our proposition that **sanctions in relation to expectation violations are about specific features given app developers' drive to target specific market niches.**

*Table 3. Prevalence of features associated with Disappointment, Complaints, Turning to Competitors, Advice Seeking and Giving, and Uninstalling across the three Apps*

| Disappointment | | | Complaints to App Proprietor | | | Turn to competitors | | | Advice | | | Uninstalling | | |
|---|---|---|---|---|---|---|---|---|---|---|---|---|---|---|
| Features | Counts | % | Features | Count | % | Features | Count | % | Features | Count | % | Features | Counts | % |
| Zombies Run | 102 | 11.4 | GPS | 47 | 11.6 | GPS | 92 | 27.2 | Maps | 37 | 13.3 | GPS | 36 | 14.5 |
| GPS | 94 | 10.5 | Maps | 26 | 6.4 | Distance tracking | 29 | 8.6 | GPS | 19 | 6.8 | Privacy violation | 29 | 11.7 |
| Recording | 60 | 6.7 | Distance tracking | 23 | 5.7 | Recording | 28 | 8.3 | Phone compatibility | 14 | 5.0 | Distance tracking | 17 | 6.9 |
| Maps | 58 | 6.5 | MyTracks | 21 | 5.2 | Maps | 19 | 5.6 | Functionality* | 12 | 4.3 | Battery* | 16 | 6.5 |
| Distance tracking | 56 | 6.3 | Privacy violation | 20 | 5.0 | Route tracking | 16 | 4.7 | Audio | 11 | 3.9 | Recording | 15 | 6.0 |
| MyTracks | 42 | 4.7 | Installing/updating | 18 | 4.5 | MyTracks | 14 | 4.1 | Recording | 11 | 3.9 | Maps | 12 | 4.8 |
| Functionality* | 32 | 3.6 | Login* | 18 | 4.5 | Phone compatibility | 12 | 3.6 | Facebook app* | 9 | 3.2 | MyTracks | 10 | 4.0 |
| Route tracking | 32 | 3.6 | Customer service* | 15 | 3.7 | Speed tracking* | 9 | 2.7 | Distance tracking | 8 | 2.9 | Endomondo* | 9 | 3.6 |
| Installing/updating | 31 | 3.5 | Battery* | 14 | 3.5 | Privacy violation | 8 | 2.4 | Route tracking | 8 | 2.9 | Zombies Run | 9 | 3.6 |
| Audio | 28 | 3.1 | Recording | 14 | 3.5 | Calorie counter* | 7 | 2.1 | Zombie Infestation* | 8 | 2.9 | Route tracking | 6 | 2.4 |
| | | | Route tracking | 14 | 3.5 | | | | | | | Phone compatibility | 6 | 2.4 |

**Note**. * = feature was excluded from the chi-squared tests due to small expectancy counts or because they were not associated with a particular sanction.

*Table 4. Prevalence of issues associated with Disappointment, Complaints, Turning to Competitors, Advice Seeking and Giving, and Uninstalling across the three apps*

| Disappointment | | | Complaints to App proprietor | | | Turn to competitors | | | Advice | | | Uninstalling | | |
|---|---|---|---|---|---|---|---|---|---|---|---|---|---|---|
| Issues | Count | % | Issues | Count | % | Issues | Count | % | Issues | Count | % | Issues | Count | % |
| Bad/boring* | 133 | 14.9 | Inaccurate tracking | 52 | 12.9 | Inaccurate tracking | 63 | 18.6 | Inaccurate tracking | 19 | 6.8 | Inaccurate tracking | 33 | 13.3 |
| Inaccurate Tracking* | 103 | 11.5 | Access to private data | 18 | 4.5 | GPS signal loss | 48 | 14.2 | Can't share tracks | 13 | 4.7 | Access to private data | 27 | 10.9 |
| Does not work | 39 | 4.4 | Customer support unhelpful | 14 | 3.5 | No location fix | 14 | 4.1 | Couldn't figure out how to play | 12 | 4.3 | Drains battery | 17 | 6.9 |
| GPS signal loss* | 38 | 4.3 | Drains battery | 14 | 3.5 | Bad/boring | 11 | 3.3 | Can't kill zombies | 11 | 3.9 | GPS signal loss | 15 | 6.0 |
| Force closes/crashes | 29 | 3.3 | Bad/boring | 14 | 3.5 | Access to private data | 8 | 2.4 | Can't send tracks to maps | 9 | 3.2 | Bad/boring | 12 | 4.8 |
| No audio | 26 | 2.9 | Can't install – package | 13 | 3.2 | Force closes/crashes | 7 | 2.1 | GPS signal loss | 8 | 2.9 | No location fix | 8 | 3.2 |
| No location fix | 23 | 2.6 | GPS signal loss | 12 | 3.0 | Cuts out sections of route | 6 | 1.8 | No audio | 7 | 2.5 | Force closes/crashes | 8 | 3.2 |
| Can't track distance | 17 | 1.9 | Does not work | 12 | 3.0 | | | | No support for HRM | 7 | 2.5 | Interferes with phone functions | 6 | 2.4 |
| Phone incompatibility | 16 | 1.8 | Can't share tracks | 11 | 2.7 | | | | Drains battery | 6 | 2.2 | Cuts out sections of route | 6 | 2.4 |
| Cuts out sections of route | 15 | 1.7 | No location fix | 9 | 2.2 | | | | | | | | | |

**Note**. * = Only three features were included in the chi-squared tests as the remaining 17 were eliminated due to small expectancy counts or because they were not associated with a particular sanction.

### *4.3. Similar sanctions to expectation violations (H3)*

To test whether users of a specific app domain will respond with similar sanctions, we calculated the prevalence of sanctions separately for MyTracks ($f_{\text{My Tracks}}$ = 984), Endomondo ($f_{\text{Endomondo}}$ = 910) and Zombies Run! ($f_{\text{Zombies Run!}}$ = 608). The prevalence of sanctions is presented below in Table 5. Here it is seen that both MyTracks and Endomondo users responded most frequently to expectation violation with expressions of disappointment, followed by complaints, turning to competitors and uninstalling. In fact, in Table 5 the pattern of sanctions is identical for the two apps with the exception of the payment complaints for Endomondo. Apart from disappointment, which accounted for 50% of Zombie Run! users' sanctions, the overall pattern of sanctions for Zombie Run! was not consistent with those for MyTracks and Endomondo. Table 5 shows that, beyond disappointment sanctions, users responded more frequently to expectation violation by warning other users more than MyTracks and Endomondo users. Interestingly, MyTracks users did not make any payment complaints (e.g., about cost and membership subscription), although, all three apps offer in-app purchases. Further, only Zombies Run! users responded to expectation violation by blaming themselves and expressing embarrassment. The prevalence of sanctions (for the seven sanctions with non-empty values in Table 5) is also shown in Figure 4.

*Table 5. Prevalence of sanctions in response to expectation violation for the three apps*

| **Sanction** | **MyTracks** | | **Endomondo** | | **Zombies Run!** | |
|---|---|---|---|---|---|---|
| | Count | % | Count | % | Count | % |
| Disappointment | 355 | 36.1 | 233 | 25.6 | 304 | 50.0 |
| Complaints | 167 | 17.0 | 184 | 20.2 | 53 | 8.7 |
| Turn to competitors | 152 | 15.4 | 176 | 19.3 | 10 | 1.6 |
| Advice giving or seeking | 109 | 11.1 | 87 | 9.6 | 83 | 13.7 |
| Uninstalling | 118 | 12.0 | 98 | 10.8 | 32 | 5.3 |
| Warning to users | 53 | 5.4 | 31 | 3.4 | 74 | 12.2 |
| Payment complaints | - | - | 93 | 10.2 | 7 | 1.2 |
| Abuse | 30 | 3.0 | 8 | 0.9 | 20 | 3.3 |
| User blames self | - | - | - | - | 15 | 2.5 |
| Embarrassment | - | - | - | - | 10 | 1.6 |
| **Total counts** | 984 | | 910 | | 608 | |

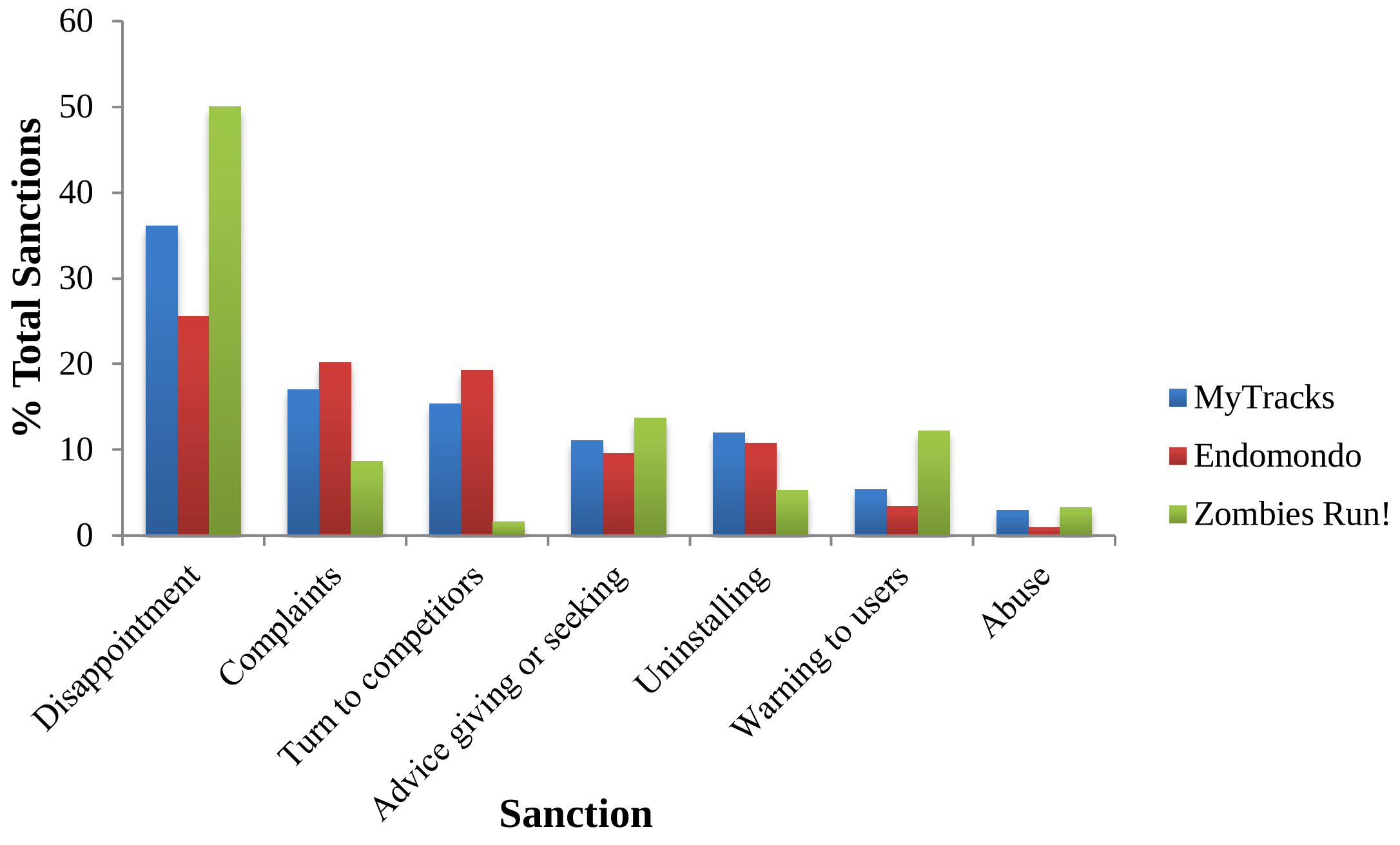

*Figure 4. Percentage of sanctions as common responses to expectation violation across MyTracks, Endomondo and Zombies Run! apps*

Analysing the sanctions prevalent across all apps, the chi-squared test revealed a significant difference in prevalence of sanctions between apps, $\chi^2(12)= 263.02$, $p < .001$, Cramer's $V = .24$. Furthermore, our outcomes revealed similarities between MyTracks and Endomondo, as shown above in Table 5 and Figure 4. Neither MyTracks or Endomondo users uninstalled their apps more often than expected by chance, *zs* = 1.5 and 1.4 respectively. They also did not differ significantly in how often they gave or sought advice, $z$ = -.6 and -.9. MyTracks and Zombies Run! users were comparable in that they did not respond with abuse beyond chance, *zs* = 1.2 and 1.6 respectively.

Disappointment was expressed less frequently towards Endomondo, $z = -4.2$, $p < .01$, but more frequently towards Zombies Run! than expected, $z = 6.0$, $p < .01$. Complaints directed towards Endomondo occurred more often than expected, $z = 3.8$, $p < .01$. However, users responded with fewer complaints towards Zombies Run!, $z = -4.5$, $p < .01$. Disappointment and complaints were not significantly associated with MyTracks beyond chance, *zs* = -7 and 0.

Similarly, users of Endomondo more frequently responded to expectation violation by turning to competitors than expected, $z = 5.6$, $p < .01$. Over 19% of turning to competitor sanctions were directed at Endomondo. Again, turning to competitors occurred less frequently for Zombies Run! users than expected, $z = -7.9$, $p < .01$. MyTracks users did not turn to competitors beyond chance, $z = 1.0$.

Contrary to MyTracks and Endomondo, Zombies Run! users responded with advice-based sanctions marginally more often than expected, $z = 1.9$, $p < .1$, but considered or decided to uninstall their apps significantly less often than expected, $z = -3.6$, $p < .01$.

Although warning to users and abuse occurred less often overall, as indicated in Table 5, Endomondo users responded to expectation violation by warning others less often than expected, $z = -3.2$, $p < .01$, but Zombies Run! users warned others more often than expected, $z = 5.8$, $p < .01$. MyTracks users did not significantly warn others beyond chance, $z = -1.5$. Finally, unlike MyTracks and Zombies Run! users, Endomondo users responded less often with abuse than expected, $z = -2.7$, p < .01.

Overall, in the consideration of H3, **the results does not fully support the argument that users of a specific app domain will respond with similar sanctions to expectation violations given their** **anticipations of how software features should work in that domain.** That said, the pattern of results in Figure 4 holds across all three apps, and outcomes were duplicated for MyTracks and Endomondo. However, there was slight variation in the outcomes for Zombies Run! (and particularly for the disappointment sanction).

We also tested whether users will have similar expectations of how features work within specific app domains. Consistent with the above feature-sanction associations, we took the five most prevalent sanctions and analysed feature malfunctions commonly associated with each of the five sanctions for the three apps. Out of the 17 MyTracks features that were most often targeted for expectation violations across the five sanctions (as shown in Table 6), we eliminated 13 features with expectancy counts of less than 5 after two preliminary chi-squared tests. Thereafter, we conducted a final chi-squared test to explore the association between the top five sanctions and the following four features: GPS, maps, distance tracking and general dissatisfaction towards the app.

As shown in Table 6, we found significant differences in the prevalence of feature-sanction associations for MyTracks users, $\chi^2(12) = 65.00$, $p < .001$, Cramer's $V = .23$. Users did not express disappointment towards any of the features significantly more often than expected, *zs* range = -1.2 to 1.2. Similarly, users did not respond to feature malfunctions significantly more often than expected by complaining, *zs* range = -.9 to .9. However, users often responded to GPS malfunctions, $z = 2.9$, $p < .01$ by turning to competitors. General dissatisfaction with the MyTracks app and distance tracking did not cause users to turn to competitors more often than expected, *zs* =-1.4 and -.5. In contrast, users responded less often with advice-based sanctions to GPS malfunctions, $z = -2.1$, $p < .05$, but significantly more often than expected to map malfunctions, $z = 5.5$, $p < .01$. Finally, uninstalling was not associated with any of the features above chance expectancy, *zs* range = -.9 to .4.

For Endomondo, we used the same feature selection procedure as above. While there were 23 different features expressed frequently across the five sanctions (see Table 7), sanctions were only expressed consistently towards three specific features: recording, GPS and distance tracking. However, there were no significant differences in the prevalence of feature-sanction associations, $\chi^2(8) = 10.38$, $p > .1$.

Similarly, when applying the same process of analysis to the sanctions recorded for Zombies Run! features our outcomes were different (refer to Table 8). We observed that users of Zombies Run! directed different sanctions towards a wide range of features. However, GPS, maps and general dissatisfaction with the app were the more prevalent features. Of note here is that for all three apps, GPS-related violations and sanctions were consistent, in support of our pattern of outcomes in Figure 4.

*Table 6. Prevalence of MyTracks features associated with frequent sanctions*

| Disappointment | | | Complaints to App proprietor | | | Turn to competitors | | | Advice | | | Uninstalling | | |
|---|---|---|---|---|---|---|---|---|---|---|---|---|---|---|
| Features | Count | % | Features | Count | % | Features | Count | % | Features | Count | % | Features | Counts | % |
| GPS | 49 | 13.8 | GPS | 27 | 16.2 | GPS | 50 | 32.9 | Maps | 29 | 26.6 | Privacy violation | 26 | 22 |
| MyTracks | 42 | 11.8 | MyTracks | 21 | 12.6 | My Tracks | 14 | 9.2 | GPS | 8 | 7.3 | GPS | 16 | 13.6 |
| Maps | 32 | 9.0 | Maps | 19 | 11.4 | Maps | 13 | 8.6 | Recording | 7 | 6.4 | Battery | 14 | 11.9 |
| Distance tracking | 29 | 8.2 | Installing/updating | 13 | 7.8 | Distance tracking | 10 | 6.6 | File imports | 5 | 4.6 | MyTracks | 10 | 8.5 |
| Route tracking | 24 | 6.8 | Privacy violation | 11 | 6.6 | Recording | 10 | 6.6 | App compatibility | 4 | 3.7 | Maps | 7 | 5.9 |
| Installing/updating | 22 | 6.2 | Route tracking | 10 | 6.0 | Route tracking | 7 | 4.6 | MyTracks | 4 | 3.7 | Distance tracking | 5 | 4.2 |
| Recording | 19 | 5.4 | Elevation tracking | 8 | 4.8 | User interface | 5 | 3.3 | Route tracking | 4 | 3.7 | Recording | 5 | 4.2 |
| Elevation tracking | 15 | 4.2 | Distance tracking | 7 | 4.2 | Privacy violation | 4 | 2.6 | Widget | 4 | 3.7 | Installing/updating | 5 | 4.2 |
| Speed tracking | 12 | 3.4 | Battery | 7 | 4.2 | | | | Heart rate monitor | 3 | 2.8 | Route tracking | 4 | 3.4 |
| User interface | 12 | 3.4 | User interface | 6 | 3.6 | | | | | | | Statistics | 4 | 3.4 |

*Table 7. Prevalence of Endomondo features associated with frequent sanctions*

| Disappointment | | | Complaints to App proprietor | | | Turn to competitors | | | Advice | | | Uninstalling | | |
|---|---|---|---|---|---|---|---|---|---|---|---|---|---|---|
| Features | Count | % | Features | Count | % | Features | Count | % | Features | Count | % | Features | Count | % |
| Recording | 38 | 16.3 | Login | 18 | 9.8 | GPS | 38 | 21.6 | Facebook app | 9 | 10.3 | GPS | 12 | 12.2 |
| GPS | 36 | 15.5 | GPS | 18 | 9.8 | Distance tracking | 19 | 10.8 | GPS | 7 | 8.0 | Distance tracking | 12 | 12.2 |
| Distance tracking | 26 | 11.2 | Distance tracking | 16 | 8.7 | Recording | 16 | 9.1 | Distance tracking | 6 | 6.9 | Recording | 10 | 10.2 |
| Endomondo | 14 | 6.0 | Facebook app | 13 | 7.1 | Route tracking | 9 | 5.1 | App compatibility | 6 | 6.9 | Endomondo | 9 | 9.2 |
| Maps | 9 | 3.9 | Customer service | 12 | 6.5 | App compatibility | 8 | 4.5 | Audio | 5 | 5.7 | Maps | 5 | 5.1 |
| Login | 8 | 3.4 | Endomondo | 9 | 4.9 | Customer service | 6 | 3.4 | Maps | 4 | 4.6 | Stability/ functions | 5 | 5.1 |
| Activities | 7 | 3.0 | Privacy violation | 9 | 4.9 | Login | 6 | 3.4 | Recording | 4 | 4.6 | Subscription service | 4 | 4.1 |
| Audio | 7 | 3.0 | Recording | 9 | 4.9 | Speed tracking | 6 | 3.4 | Route tracking | 4 | 4.6 | Advertisements | 3 | 3.1 |
| Calorie counter | 6 | 2.6 | Battery | 7 | 3.8 | Maps | 5 | 2.8 | Phone compatibility | 4 | 4.6 | App compatibility | 3 | 3.1 |
| Installing/ updating | 5 | 2.1 | Subscription service | 6 | 3.3 | Subscription service | 5 | 2.8 | Activities | 3 | 3.4 | Music player | 3 | 3.1 |
| Route tracking | 5 | 2.1 | | | | | | | Track editing | 3 | 3.4 | Privacy violation | 3 | 3.1 |
| Speed tracking | 5 | 2.1 | | | | | | | | | | | | |

*Table 8. Prevalence of Zombies Run! features associated with frequent sanctions*

| Disappointment | | | Complaints to app proprietor | | | Turn to competitors | | | Advice | | | Uninstalling | | |
|---|---|---|---|---|---|---|---|---|---|---|---|---|---|---|
| Features | Count | % | Features | Count | % | Features | Count | % | Features | Count | % | Features | Count | % |
| Zombies Run | 102 | 33.6 | Functionality | 9 | 17.0 | Competition | 1 | 10.0 | Functionality | 11 | 13.3 | Zombies Run | 9 | 28.1 |
| Functionality | 29 | 9.5 | Zombies | 6 | 11.3 | Entertainment value | 1 | 10.0 | Zombie infestation | 8 | 9.6 | GPS | 7 | 21.9 |
| Audio | 21 | 6.9 | Phone compatibility | 4 | 7.5 | GPS | 4 | 40.0 | Wrong version | 7 | 8.4 | Functionality | 4 | 12.5 |
| Maps | 17 | 5.6 | Zombies Run | 4 | 7.5 | Location | 1 | 10.0 | Audio | 6 | 7.2 | Wrong version | 2 | 6.3 |
| Phone compatibility | 15 | 4.9 | Audio | *3* | 5.7 | Maps | 1 | 10.0 | Competition | 5 | 6.0 | Zombies | 2 | 6.3 |
| Zombies | 13 | 4.3 | Error message | 3 | 5.7 | Requests detours | 1 | 10.0 | Options | 5 | 6.0 | Help | 2 | 6.3 |
| Location | 12 | 3.9 | Features | 2 | 3.8 | Zombies Run | 1 | 10.0 | GPS | 4 | 4.8 | Error message | 1 | 3.1 |
| GPS | 9 | 3.0 | Game | 2 | 3.8 | | | | Location | 4 | 4.8 | Features | 1 | 3.1 |
| Movement | 9 | 3.0 | GPS | 2 | 3.8 | | | | Maps | 4 | 4.8 | Game | 1 | 3.1 |
| Error message | 5 | 1.6 | Maps | 2 | 3.8 | | | | Instructions | 3 | 3.6 | Request detours | 1 | 3.1 |
| Force closes/crashes | 5 | 1.6 | Movement | 2 | 3.8 | | | | Weapons | 3 | 3.6 | Websites | 1 | 3.1 |
| | | | Uninstalling | 2 | 3.8 | | | | Zombies | 3 | 3.6 | | | |

## 5. Discussion and Implications

The aim of this study was to explore expectation violation in mobile apps. In Section 2 we explored Expectancy Violation Theory [1, 2, 15] and Expectation Confirmation Theory [16], and the way users must increasingly interact with software in contemporary societies. Such interactions give rise to likely expectations about how various types of software 'should behave', as seen in a human-human context. Thus, we anticipated that *a mobile app community responds to expectation violations with sanctions in their reviews when features do not work as anticipated (H1)*. In addition, with the ease of entering the mobile app marketplace and the many developers releasing apps, the need to be nimble means that apps are developed with the view of perfecting the same features of other apps in the domain [20, 47, 48], referred to as a "non-additivity" strategy among those that considered the demand theory [22]. From a marketing perspective, this trend follows Porter's strategy aimed at market differentiation, where products and services are qualified based on their superiority [23].

To this end, we proposed that *sanctions in relation to expectation violations are about specific features given app developers' drive to target specific market niches (H2)*. More specifically, sanctions directed towards specific features shed light on users' prior expectations of how an app should work [41, 42]. Furthermore, generalised expectations tend to influence formal rules or laws in normal society [24], which may also extend to the way users develop expectations in socio-technical systems, and mobile apps in particular. We thus articulated that *users of a specific app domain will respond with similar sanctions to expectation violations given their anticipations of how software features should work in that domain (H3)*. Testing these hypotheses would reveal key insights into human-software interactions in socio-technical systems, which would provide knowledge for the information systems community.

To conduct our investigation aimed at testing these three hypotheses we explored the prevalence of sanctions and the strength of their associations with features and issues (problems) across three popular health and fitness apps, MyTracks, Endomondo and Zombies Run!. Here we revisit our outcomes, discussing our findings and considering their implications. We consider each hypothesis in turn in the following three subsections (Sections 5.1 – 5.3).

### *5.1. App community responds to expectation violations with sanctions (H1)*

Collapsing across apps, we found that users responded to expectation violation with different sanctions, which supported H1. In particular, users responded more often to expectation violation by expressing disappointment, making complaints and turning to competitors, which indicated that, on the whole, they preferred to adopt more active approaches over passive approaches (e.g., giving or seeking advice and warning other users as against just quietly refraining for using apps), in response to expectation violation. However, in the present study we only explored expectation violation of the five most prevalent sanctions, which accounted for nearly 87% of the total number of sanctions, and were most strongly associated with feature malfunctions. Our results are consistent with ECT [15] and EVT, in that consumers have preconceived notions and expectations about how fitness apps should work [41, 42] These expectations are often based on prior experience, social schemas or unstated rules [9, 88]. The results also suggest that users respond to expectation violation of apps through a limited number of sanctions which are specific to the convention or norms prescribed by the app domains [8, 33]

Our evidence confirms the relevance of EVT and ECT for studying expectation violations in socio-technical systems, beyond human-human applications [3, 4, 6, 7]. EVT notes that individuals form expectations based on societal rules which drives their beliefs about future events, and when the expectation of an individual is violated they react in compensation [8, 33]. In this work we observe evidence of a range of sanctions, with the most severe being "turning to competitors", "uninstalling" and "warning other users" – all associated with strong dissatisfaction [37, 38]. ECT notes that customer fulfilment is influenced by previous expectations, perceived performance, and disconfirmation of beliefs. In the mobile app context, a socio-technical system or virtual community, external behavioural change (i.e., a verbal sanction) is fitting, which could no doubt result in loss of reputation [1, 2], and particularly with users' drive to warn other users when their expectations were violated. Since there are limited if any opportunities for users to meet with app providers, users may freely express their feelings online when frustrated, which provokes the thought - *perhaps virtual communities promote greater willingness to express verbal sanctions when expectations are violated?* Confirming such a finding with follow up work could have implications for theory, leading for instance to contextualisation of EVT and ECT. In fact, our outcomes here somewhat extends ECT. While this theory notes that previous expectations, perceived performance, and disconfirmation of beliefs impacts customer fulfilment, here we observe that the lack of mobile app users' fulfilment results in sanctions.

### *5.2. Expectation violations about specific features (H2)*

In addition to the distinct patterns of sanctions, we also found significant differences in the associations between sanctions and particular feature malfunctions, which supported H2. Users responded to general Zombies Run! malfunctions more often than expected with expressions of disappointment, but less often with complaints and advice and by turning to competitors. However, GPS was the most prominent individual feature that elicited sanctions when the user experienced malfunctions. Because an accurate GPS is a vital feature of many fitness apps, users have strict guidelines and schemas about how it should operate (e.g., [14]). Thus, norm violation leads users to turn to competitors more often and express disappointment less. Similar associations were found between strong, active sanctions (e.g., uninstalling and complaint) and privacy violation.

Privacy violation is a common issue and concern of individuals who use computer-mediated communication systems, such as social networking sites (e.g., [89, 90]), e-commerce (e.g., [91, 92]) and even blog services [93]. According to Social Contract Theory [94], the e-commerce user may consider a company's access to personal information to be fair if they have control over the information sharing process [95] and is informed about how their information will be used. It is also important that the user and company share an understanding of the rules of their contract [92]. Malhotra et al. [92] found that e-commerce requests for more sensitive personal information from users reduces both trust in the company and continuance intention. In the present study, the response of fitness app users was more extreme as they expressed they did not feel it was appropriate for apps to extract any personal information (such as phone contacts). Thus, users responded to privacy violation less often with expressions of disappointment and more often with complaints and by uninstalling their fitness app. In other words, fitness app users determined whether or not an app was worthwhile based on whether they experienced privacy violation and discontinued use based on this occurrence.

We also found similar patterns of active versus passive sanctions resulting from specific feature-based issues. Although the results still supported H2, sanctions were not directed towards as many specific issues to the same extent as feature malfunctions. One of the most prevalent issues was the users' perceptions that their apps were bad/boring: a superficial characteristic of the app that is not vital to its functioning and level of usability. Perception that the app is bad/boring was more often responded to with expressions of disappointment and less often with complaints, turning to competitors and advice-based sanctions. In contrast, inaccurate tracking and GPS signal loss, both of which are vital to the functioning of a fitness app, were associated with fewer expressions of disappointment as users took active approaches to resolving dissatisfaction. Inaccurate tracking more often resulted in complaints and GPS signal loss more often resulted in users turning to competitors. This is consistent with applications of the Technology Acceptance Model [96-98] revealing that perceived usefulness and satisfaction of mobile internet services are stronger predictors of continuance intention than perceived enjoyment of the superficial characteristics of the service [99].

Our outcomes also confirm that with the ease of entering the mobile app marketplace and the many developers releasing app, there seems to be a drive towards a strategy aimed at market differentiation where products and services are qualified based on their superiority [23]. While such a strategy may find favour among users that are driven by features they are familiar with, especially having used similar apps that performed unsatisfactorily, challenges still remains for app developers in terms of addressing users' needs. This is particularly essential as users seem to freely express their feelings online when frustrated, and lack of fulfilment will likely result in sanctions.

### *5.3. Similar sanctions to expectation violations (H3)*

For the most part, users tended to respond with similar sanctions to expectation violations for apps in a given app domains. However, our outcomes in the previously section do not fully support H3. In particular, Zombies Run! users responded to expectation violation with a distinctive pattern of sanctions. Users of Zombies Run! expressed more disappointment, warned other users more and made marginally more advice-based sanctions than MyTracks and Endomondo users. However, they were significantly less likely to direct their complaints at the app proprietors, turn to competitors and uninstall their app after expectation violation. They also exhibited unique consumer behaviours such as blaming themselves and expressing embarrassment as a consequence of their dissatisfaction with the app. Although self-blaming did not occur frequently overall, it repeatedly co-occurred with difficulty figuring out how to play. These patterns of sanctions suggest that Zombies Run!, which operates like a simple tablet game, is sufficiently novel as a fitness app that users have more malleable schemas and less strict expectations about how this app should function. Alternatively, users of Zombies Run! may still be dissatisfied with the app but make a rational, calculative commitment to seek compensation or simply get a reaction from the app proprietors to correct any problems (e.g., [100]), and are likely to continue using the app regardless due to switching costs and lack of viable alternatives within the same market niche (see [101, 102]).

While MyTracks and Endomondo contain gamification elements, such as competition between friends and level of achievement [103], Zombies Run! has more *game* specific elements (see [104]), such as feedback when the zombies attacked, a basic narrative context or story (i.e., zombie infestation) as well as levels of achievement that the user can reach. The results from the present study suggest that Zombies Run! users' expectations of enjoyment and engagement as a game, rather than quality of a fitness app, were violated. Surprisingly, several users of Zombies Run! also complained that they had to run to play, which implies that app developers are targeting a slightly different market niche to MyTracks and Endomondo. This is consistent with the finding that gamification elements of fitness apps were associated with users' psychological and physical capability (e.g., self-monitoring), motivation (e.g., goal-setting) and opportunity/triggers (e.g., sociability, cues to action, stimulus control), whereas game based elements were only associated with opportunity/triggers [105].

In contrast to Zombies Run!, users of Endomondo had more specific expectations of how their app should work and therefore directed their dissatisfaction towards the app proprietors; they responded to expectation violation by more often complaining and turning to competitors and warned other users less often. However, these sanctions were expressed equally across the different features when malfunctions occurred, which may suggest that the vital features are mostly reliable. The sanctions may, instead, reveal broader issues or negative user experiences.

Interestingly, users of MyTracks did not express dissatisfaction through particular sanctions beyond that by chance, which may suggest that users are, on average, satisfied with how MyTracks works or they did not feel the need to take drastic action to address expectation violation. For instance, no MyTracks feature malfunctions were associated with uninstalling beyond chance expectancy. Map malfunctions did not elicit more complaints or even disappointment, instead, users more often gave advice to and sought advice from the app proprietors. However, when GPS malfunctions did occur, users were more likely to turn to competitors and less likely to respond passively by giving advice, as this feature is vital to the functioning of MyTracks.

Our findings suggest that, as with how individuals may develop specific expectations of others given certain norms (e.g., culture-specific norms and conventions [27, 34, 35]), app users also grow to possess specific expectations and beliefs about the features that are delivered in apps given their purpose. Outcomes here also confirm that violations of expectations may be reported in a range of media outlets, including app review portals [39, 40]. Since the app domain at hand largely concerns the tracking of exercise (e.g., walks and runs) it is understandable that users of all three apps used specific features to support their exercise regime (e.g., GPS to identify their location over time). Thus, when such features did not work as expected, users complained, expressed resentment and promised reprisal. Thus, in the targeting of a strategy aimed at market differentiation, where products and services are qualified based on their superiority [23], providers in the app community should work to ensure that stringent usability evaluations are performed prior to product release in view of delivering on such promises.

### *5.4. Practical and Theoretical Implications*

This study has important implications for how app proprietors should release their products to target specific niches. App proprietors should attend to the types and frequency of sanctions that app users make when specific features or a collection of features do not meet their expectations, as this provides a strong indication of the role and niche that the app fulfils. These sanctions should also inform app proprietors of the features and resulting issues (problems) that need to be dealt with and that are considered vital to the functioning of the app.

This study further emphasized the importance of online forums and electronic word-of-mouth (e-WOM) in app development. They allow consumers to acquire and exchange information about the quality and usability of particular products (e.g., [106, 107]). However, previous research has focussed more on service efficiency, such as fulfilment and system availability [108], service failure in online retailing (e.g., [100, 109, 110]), and personal motives as antecedents of complaining behaviour [111] rather than the expectation violation of apps in specific market niches and the types of complaints themselves. Also, unlike previous e-WOM research that broadly categorized all expressions of disappointment and resentment as complaints (e.g., [8, 33, 112]), the present study extend previous research by revealing differences in the prevalence of sanctions as a result of expectation violation.

However, despite the relevance and applicability of the present study to the field of user research, our outcomes need to be evaluated in the given context. For instance, we were limited by the number of different issue-sanction associations we could explore as some issues were evenly distributed across sanctions (as indicated by multiple issues with similarly small counts). This is surprising as one would expect issues to be consequences of particular feature malfunctions and be equally prevalent across sanctions. However, the issues were less specific than feature malfunctions, for instance, the broad issue of “inaccurate tracking” was reported more frequently than the type of inaccurate tracking (e.g., distance, speed and elevation). It is also plausible that one feature malfunction may result in multiple issues. In future, rather than analysing features and issues separately, social network analysis (SNA) could be used to explore closeness of feature-issue associations for sanctions and across app domains. SNA has previously been used to support such pattern identification requirement among related items [81].

In addition, although customer reviews were treated as our unit of analysis, we could not control for nested hierarchies in the data. Because the reviews were anonymous, it was unclear whether customers contributed more than one review each and thus within-subjects correlations across responses were not controlled for. On a similar note, we did not measure changes in the prevalence of sanctions, features and issues with the passage of time. For instance, we observed a cluster of expressions of disappointment and complaints from users who couldn’t install their app because of an error message involving a package signature. These sanctions indicate the current problems app developers face and whether they are resolved in the short or long-term. This offers a useful area for future research. Furthermore, the coding of features, issues and sanctions could be refined in the future, especially for uninstalling sanctions, which could be further categorized as “threats to uninstall” and “decisions to uninstall”, and advice-based sanctions, which could be broken down into “giving advice” and “seeking advice”, the latter of which was more prominent in Zombies Run! This provides another avenue for future research.

Finally, most of the expressed sanctions were associated with monovalent dissatisfiers, i.e., features of an app that users expect will work [16]. Considering our subset of reviews, users rarely make compliments and express satisfaction towards, for instance, GPS, recording and location fix when the features do work but nearly always express dissatisfaction when they do not. Some of our data were incorrectly coded as expressing resentments in relation to expectation violations, when in fact these were positive sentiments which could shed light on this issue. Future research should also explore monovalent satisfiers (entertainment and aesthetic features that usually only result in satisfaction) and bivalent satisfiers (features that lead to both satisfaction and dissatisfaction) as antecedents of sanctions [16, 109].

## 6. Threats to Validity

While we have provided a range of insights in this work, we acknowledge that there are several shortcomings that may potentially affect the generalisability of our study outcomes; we consider these in turn.

Our work only considered review sentences that were rated =< 3 to contain potential responses to expectation violations. While previous work has indicated that such reviews largely contain expressions of dissatisfaction [66], we concede that reviews rated 4 and 5 may also contain expectation violations which are not considered in this work.

We acknowledge that we may have missed some negative words, and particularly those that may be specific to app users, but not captured as a part of the LIWC library or our negative modal verbs and common slangs list. In addition, there is a possibility that our analysis comprised of false positives due to the way different individuals use phrases. That said, our conventional inductive content analysis unearthed 446 false positive sanctions, and we did not observe many unconventional negative words use. Thus, incidence of such words and false positives were captured by our analysis.

We are aware that the reviews that were analysed in this work do not necessarily represent those of all users of the apps considered, and furthermore, some introverted users may have sentiments but suppress them altogether. Thus, our findings do not necessarily provide a complete picture of app users' views. That said, we believe that our approach and the insights that are provided could be valuable to stakeholders of the app community in general.

Our content analysis involving interpretation of textual data is subjective, and so questions may naturally arise regarding the validity and reliability of the outcomes of this analysis. In this work we conducted several rounds of reliability checks, with our inter-rater outcomes indicative of excellent agreement [77]. However, we also recognise that additional research employing interview techniques would serve to further validate the outcomes of these analyses. While our work looked to make inferences largely from users' reviews, we are aware that insights into actual users' perceptions would usefully complement and extend our findings.

We studied reviews from three apps (MyTracks, Endomondo and Zombies Run!) focused on health and fitness. Users of such apps will have specific expectations and view about how such apps should work that may not apply to other app domains. What is of interest, however, was the incidence of complaints to expectation violations and sanctions in app reviews provided by the app community, and we believe that our objectives were met. That said, we believe that our results may be applicable to other apps, and particularly those in other domains, where users may express sanctions differently given certain expectations, but these would be sanctions nonetheless.

Overall then, this study shows strong rigour interpreted through the element of credibility because we provided a systematic procedure for data coding and thematic extraction that researchers can follow in the future [113]. The findings of this study also reflect high transferability, as our results have implications for how the mobile app community responds to expectation violations with sanctions when features do not work as anticipated, and the relevance of associated theories to unpack users' expression of behaviours. In addition, we examine how sanctions may grow in relation to inadequacies of specific features corresponding to the targeting of specific market niches, and how users of specific apps in a domain develop precise expectations around how particular features should be presented. We anticipate that the approaches used in this work may be successfully implemented by other researchers (in other domains) to reveal similar pattern in users' data.

## 7. Conclusion

With a shift towards the use of information technology and software services for most aspects of contemporary life (e.g., for military, commerce, agriculture, finance, education, communications, governments, private organisations, business bodies and the medical sector), there are also growing expectations about how these systems should behave, giving rise to the need to understand expectations is socio-technical systems. We have set out an agenda towards providing initial insights into expectation violations in mobile apps, by developing three hypotheses for testing (H1-H3).

We first tested whether users of the mobile app community responds to expectation violations with sanctions in their reviews when features did not work as anticipated, confirming that indeed users of apps responded to expectations with sanctions just as individuals react with compensations for the violation with sanctions in a human-human context. Our second hypothesis aimed at testing the outcomes for validating if sanctions in relation to expectation violations are about specific features given app developers drive to target specific market niches were also confirmatory. Thirdly, our findings somewhat supported our third hypothesis that users of a specific app domain respond with similar sanctions to expectation violations given their anticipations of how software features should work in that domain. These findings confirm the relevance of expectation theories for understanding commonly anticipated norms and expectations around how software systems (mobile apps) should work and the penalties and sanctions that are evoked for violating these expectations. We have also explored various theories that may provide a basis for mitigating the consequences of violations and sanctions, and explored potential implications for our outcomes. The next step in our portfolio of work is to further probe our outcomes with other techniques and theoretical lens, and perform temporal analysis.